\documentstyle[prl,aps]{revtex}
\begin{document}
\draft
\twocolumn[
\preprint{To be submitted to PRL}
\title{
Morphological Instabilities in a growing Yeast Colony:\\ 
Experiment and Theory}
\author{Thomas Sams, Kim Sneppen, and Mogens H. Jensen${^{\hbox{\ddag}}}$}
\address{Niels Bohr Institute 
  and NORDITA, Blegdamsvej 17, DK-2100 {\O}, Denmark
}
\author{Bjørn Eggert Christensen}
\address{Novo Nordisk, Novo All{\'e} 1, DK-2880 Bagsv{\ae}rd, Denmark}
\author{and}
\author{Ulf Thrane}
\address{Dept.\ of Biotechnology, Technical University of Denmark, 
     DK-2800 Lyngby, Denmark
}
\date{Nov 25, 1996}
\maketitle
\widetext
\begin{abstract}
We study the growth of colonies of the yeast {\em Pichia membranaefaciens} 
on agarose film. The growth conditions are controlled in a setup
where nutrients are supplied through an agarose film suspended over a 
solution of nutrients. 
As the thickness of the agarose film is varied, the morphology 
of the front of the colony changes. The growth of the front is modeled 
by coupling it to a diffusive field of inhibitory metabolites.
Qualitative agreement with experiments suggests 
that such a coupling is responsible for the
observed instability of the front.
\end{abstract}
\pacs{PACS 83.80.L, 87.10, 83.50.L, 68.10.C, 68.45, 64.60.F}
]
\narrowtext
The physical laws governing the development
of biological morphologies have been under investigations
at least since Turing described pattern formation 
using interacting diffusive fields \cite{Turing}.
Turing patterns have found application in a variety of 
biological systems \cite{Lee96,Goldstein96}. 
Lately the interplay between form and formation in biological
morphologies have been investigated by physical modelling 
in swarms of swimming bacteria \cite{Kessler89,Hill89}, 
in growth of bacteria colonies 
\cite{Shapiro91,Shapiro95,Wakita94,Shimada95,Ben-Jacob94a,Ben-Jacob94b},
and in growth of various fungi 
\cite{Crawford93,Patankar93,Soddell94}.

From a morphological point of view, the growth of fungi can 
be naturally divided into filamentous growth and yeast-like growth. 
Filamentous growth is characterized by a hyphal, multi-cellular, 
growth zone that under unfavorable growth conditions exhibit fractal-like 
properties \cite{Matsuura92,Matsuura93}. 
In contrast, yeasts are characterized by unicellular growth and compact 
colonies \cite{Carlile94}. 
In the present letter, the morphological properties 
of the front of yeast colonies are studied. 

In the experiment, the yeast is grown on YPD with full strength 
defined as 
10\,g/l Yeast Extract (Difco), 
20\,g/l Bacto Peptone (Difco), 
20\,g/l Dextrose, pH unadjusted. 
Agarose (Sigma type II, medium EEO) is used for solidified media. 
The experiment is carried out by streak-inoculating with cells 
of the yeast {\em Pichia membranaefaciens} \cite{A321-95} 
on the agarose film and monitoring the growth with a video camera. 
The temperature is held at 
$T=30^\circ{\rm C}$ in a chamber with relative humidity 
close to $100\,\%$. The growth is monitored with a video camera. 

In one experiment the yeast is grown on solid 0.5 strength YPD solidified 
with 0.8\,\% agarose. 
Under these conditions the growth slows down during the 
experiment and virtually stops after about 4 days. 
Examples of fronts identified from digitized images of the growing 
culture are shown in fig.\  \ref{fig:1}a. 

In a different setup, 
illustrated in fig.\  \ref{fig:experiment1}, 
the yeast is grown on a 0.3 - 2\,mm agarose film suspended 
on a Millipore TCTP14250 membrane (with 10\,$\mu$m pores) 
over continuously replenished and stirred YPD solution. 
At the membrane, an equilibrium between metabolites and nutrients 
in the reservoir and in the agarose film is established. 
In this case no retardation of the growth is observed. 
Time series of identified fronts are 
shown in figs.\  \ref{fig:1}b and \ref{fig:1}c. 

The typical size of a 
single yeast cell is $3\times 5\,\mu{\rm m}$. The
yeast grows on the agarose film with a typical doubling time of
1.6\,h. The velocity of the front is in the range
0.3 - 3\,mm/h, depending of the thickness of the 
agarose film. This means that a band of 200 - 2000 
cells contributes to the growth of the front. In the spirit of 
Trinci, we call this band the optimal growth 
zone \cite{Trinci71}.

Metabolites formed by the yeast during the degradation of nutrients 
may contain inhibitory elements. 
This is in accordance with our experimental observations: 
when there is no reservoir, the characteristic time for a metabolite to
disappear is long and a high concentration of metabolites builds up. 
This is the situation shown in fig.\  \ref{fig:1}a 
where the growth of the front is coming to a stop after a 
few days. 
The front of the colony is digitized every 5 hours and it is seen that 
the density of the growth lines increases as the velocity of the 
front slows down. 
When growing in the replenishing setup with a thin agarose film, 
the metabolites diffuse through the membrane and disappear 
on a timescale comparable to yeast growth rate.
This is the case in fig.\  \ref{fig:1}b, where the growth
does not come to a stop.
It is obvious from the fig.\  that the morphology of the front 
of the colony is very different in the two cases. 
In the case where the growth is restricted, one observes an interesting 
instability of the front 
with a cellular ``bumpy" pattern with a reasonably well-defined 
characteristic length scale. In the other case, the front of the colony 
is more compact, less ``bumpy'', and exhibits long grooves
at places where the growth for some time has been slowed down. 
This slowing down in the growth may again be connected
to the local accumulation of metabolites,
representing situations where the colony manages to grow around
an accumulation of metabolites which is then left behind in one of the grooves.
Figure \ref{fig:1}c 
shows the growth of a colony on a very thin agarose film. Here the 
growth is faster and the front of the colony  
does not show any sign of morphological instabilities.

In order to describe the observed morphological dynamics
from a theoretical point of view, 
we build upon a knowledge obtained for growth phenomena in a completely 
different environment. When an undercooled melt of a metal is subjected to a 
temperature gradient, one observes that the metal crystallizes from the 
cold side of the set-up, i.~e.\ directional 
solidification \cite{Langer80,Kertzberg83,Bechhoefer87,Sarkar87}. 
Ahead of the crystallizing front, is a field of 
impurities diffusing in the melt. The impurity field influences 
substantially the morphology of the growing front. By analogy, this is 
similar to what happens in the growth zone of the yeast colony. 
The doubling of the cells causes the appearance of the metabolites.
The metabolites can diffuse in the agarose film and 
accumulate at various places thus strongly influencing the morphology of 
the front of the colony, just like impurities will strongly influence 
the morphology of a crystallizing melt. 
To model directional solidification, one writes a 
diffusion equation of the impurities in a moving frame 
coupled to a constraint of the concentration jump over the 
front \cite{Langer80}.

Similarly, for the front of the yeast 
the formation and diffusion of the metabolites must
be an essential ingredient in the model. In terms of the 
metabolite concentration $C({\bf x},t)$, we write a diffusion 
equation on the form
\begin{equation}
\frac{d C({\bf x},t)}{dt} ~=~ D \Delta C({\bf x},t) -
 {1 \over \tau} C({\bf x},t) 
+ \theta (Y({\bf x},t))
\label{eq:1}
\end{equation}
where $D$ is a diffusion constant of the order 1\,mm${^2}$/h; 
$\tau \approx d^2/D$ is the 
characteristic time for metabolites to disappear through
the membrane where $d$ the agarose thickness, here of order $1\,{\rm mm}$. 
In the following we call $\tau$ the penetration time. 
The metabolite source term $\theta$ is arbitrarily normalized to unity 
where there is yeast present (i.e. $Y({\bf x},t)>0$).

The yeast is immersed in a liquid which it produces and carries 
along as it grows. 
The surface tension of this carrier medium introduces 
a lower cutoff $\ell$ in the possible length scales for structures of the 
front. 
The growth of the colony, which is intimately coupled to the
the concentration field $C$, is naturally averaged over
this length scale $\ell$. 
To describe the coupling
of the growth of the front to the dynamics of the suppressing metabolites
we say that the velocity of
the front depends on the concentration field by some functional
form $F(C)$, where $F$ decreases with $C$. 
When averaged over the length scale $\ell$ 
the velocity becomes
\begin{equation}
{\bf v} ({\bf x},t) ~=~ {\bf n} \,\langle F(C) \rangle_{\ell} 
 + \eta({\bf x})
\label{eq:2}
\end{equation}
where the vector ${\bf n}$ is a unit vector 
along the normal of the front and $\eta$ is a Gaussian uncorrelated 
quenched noise
with zero mean and amplitude $\Gamma$: 
$\langle \eta ({\bf x}) \eta({\bf x'}) \rangle 
= \Gamma \delta({\bf x-x'})$.
The quenched noise represents inhomogeneities in the film. 
We limit our discussion to the quenched case and do not discuss 
a possible annealed part representing temporal fluctuations in the local 
population dynamics of the yeast colony. 

The simulations are performed on a lattice. 
Each lattice site is assigned a random number $\eta({\bf x})$
between $-\Gamma$ and $\Gamma$, and a counter $w({\bf x}, t )$ 
which initially is set equal to $\eta({\bf x})$.
At each timestep we consider invasion of yeast in
all lattice sites which are nearest
neighbours to the growing colony, meaning
that the counter for all border sites are incremented by an amount:
\begin{equation}
\frac{dw({\bf x},t)}{dt} ~=~ F(C)\, n_{nn}({\bf x},t)\, S_{\ell}({\bf x},t)
\label{eq:3}
\end{equation}
When $w$ takes a value greater than $1$ 
the site is invaded, and in the next
time step new neighbours then start to be invaded.
In eq.\ (\ref{eq:3}) $n_{nn} ({\bf x})$ determines the 
number of nearest neighbours
occupied by yeast.
$S({\bf x},t)$ models the surface tension at ${\bf x}$ and is
set equal to the fraction of points within a certain 
length scale $\ell$ from ${\bf x}$ which have already been fully
occupied by the advancing colony. 
The form of the function $F(C)$ is assumed to be 
$\propto \min(1,(C_0/C)^2)$.
This functional form is an assumption; one could also
use other forms, like linear or exponential decay. 
We have used various functions in the simulations and in all cases where 
the decay of $F$ versus $C$ is not too weak
we observe qualitatively the same results as will be presented here.

In our phenomenological model 
a number of parameters enter. 
Some can be scaled out and others are
under experimental control. In summa there are 3 
important parameters:
the coupling strength to the metabolic field, determined effectively
by $F$, the penetration time $\tau$, and the strength
of the noise $\Gamma$.
In addition there is the diffusion constant $D$ which is fixed
by the size of the metabolite to be of order 1\,mm${^2}$/h
(equal that of a protein in water), and there is the
surface tension scale $\ell$
which determine our lower cut off, and is experimentally
of the size of the growth zone. 

First let us consider two extreme cases of the model.
Case one is the limit where there is no 
coupling to the metabolic field. Then the envelope of the
yeast colony will grow as a random deposition model,
exhibiting a rough surface on scales larger than
$\ell$ with statistics as the Eden growth model
in statistical physics \cite{Zhang95}. 
In case two, the coupling to the metabolic
field is very strong, and the growth will only occur at the extremely
exposed points of the envelope, and on long time scales
one observes a growth dominated by a single large ``dendrite''
eventually with side branches if simulated as here
on a quadratic lattice.

\begin{figure}
\caption[x]{
Time series from the fronts obtained from simulations of
the model defined by eqs.\  (\ref{eq:1}) and (\ref{eq:2}). 
The following parameters
are used: $D = C_0 = \Gamma = 1$, $\ell = 4$. The thickness is
varied as: (a) $d = 20$; (b) $d = 4$; (c) $d = 3$.
In all cases we simulate eqs.\  (\ref{eq:1}) and (\ref{eq:2}) a time
$t=120$ after initially seeding with 
a uniform boundary layer of yeast.
}
\label{fig:3}
\end{figure}
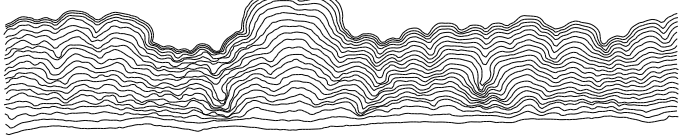
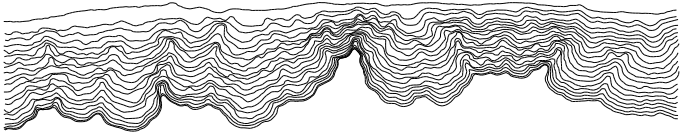
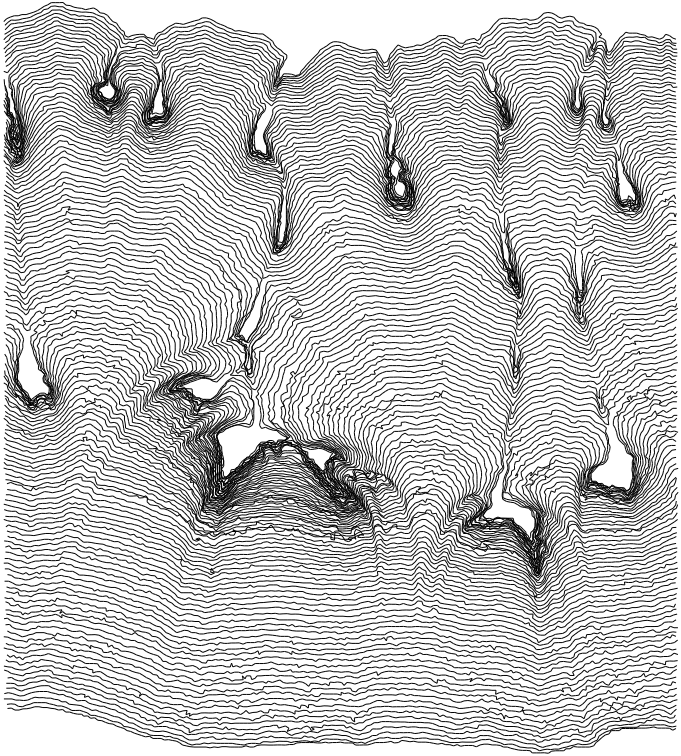
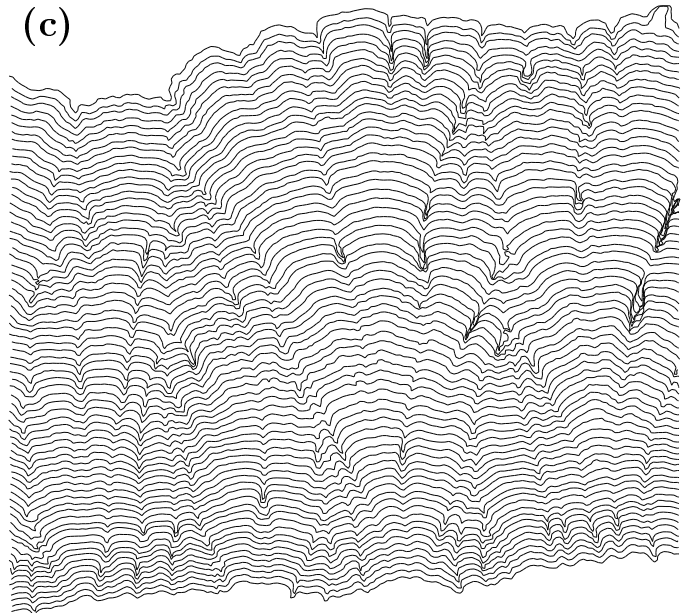
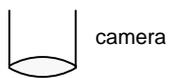
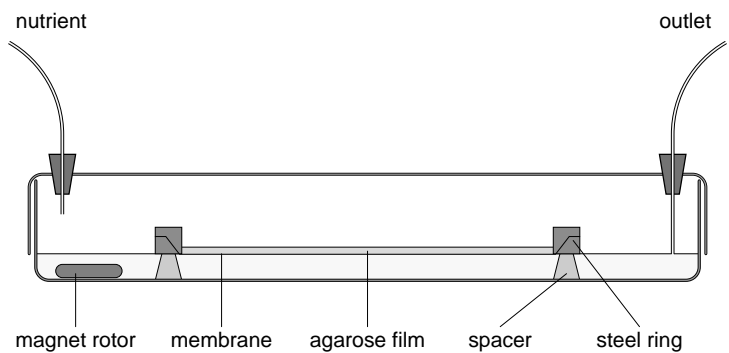
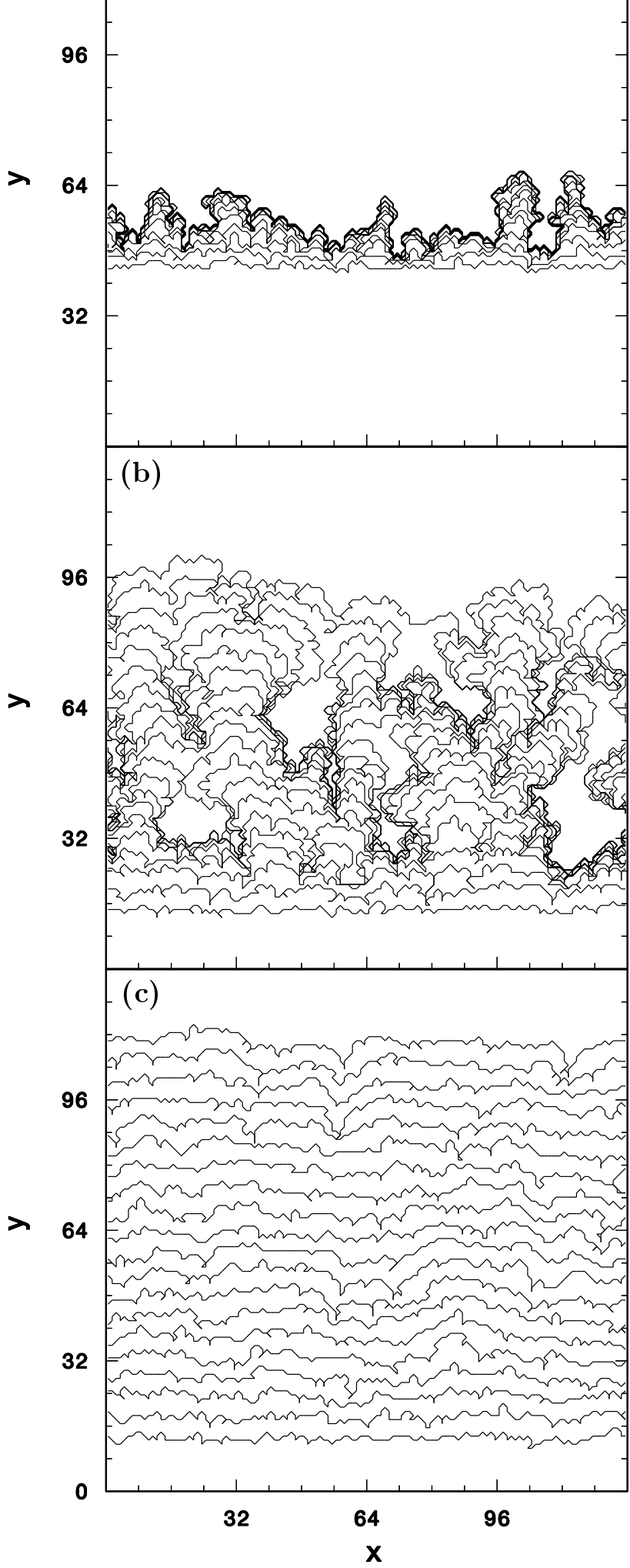

To model the experimental findings shown in fig.\  \ref{fig:1},
we investigate a 
series of situations of the model in eq.\ (\ref{eq:1}), in 
which only one of the parameters, the penetration time $\tau$ is varied. 
Figure \ref{fig:3} shows a series of growth patterns of the model 
with the following fixed values of the parameters: the diffusion 
constant is $D = 1$, the coupling to the concentration field
is $F(C)=\min(1,1/C^2)$, i.e. $C_0=1$ and
the length over which the front
is averaged $\ell=4$ (in units of the lattice spacing), 
and the noise parameter $\Gamma=1$.
We vary $d$ and set $\tau=d^2/D$ and use
$d=20$ in \ref{fig:3}a (virtually corresponding to the absence of 
the reservoir),  
$d=4$ in \ref{fig:3}b,  
and $d=3$ in \ref{fig:3}c. 
In all cases we simulate 120 timesteps
of our model, which allow a direct comparison
of absolute velocities.
In the first case with a large value of $\tau$, we observe in
fig.\  \ref{fig:3}a a front with a cellular instability, 
very much in accordance with the experiment. 
Note that there is a tendency for the larger instabilities to 
dominate over the smaller. In fact, there is a tendency 
for the characteristic length scale of the pattern
to grow slowly, in accordance with the experiment. 
What happens is that as a large instability grows forward, it 
leaves the metabolic field behind thus causing the smaller 
instabilities to be exposed to a strong metabolite field. This 
wavelength selection and screening effect is well-known in the growth of 
viscous fingers \cite{Langer80}.

In the next case, where the value of $d=4$, we observe a completely 
different morphology. In this case large grooves develop from the front 
and into the colony. These grooves are due to the fact that the 
metabolite field locally hinders the growth for a while. As time 
progresses, however, the colony grows around the place where the growth 
slowed down and the colony might then recombine again atop of the 
place with high $C$ region. There is again a good 
qualitative agreement between experiment and theory.

In the final case, where $d=3$, neither a cellular instability nor a 
complicated pattern of grooves are observed. Instead, the metabolite 
field so easily disappears through the loss term (that is through the 
membrane) that the colony grows compactly. Indeed, there will still be an 
inhomogeneity in the $C$ field causing small disturbances which develop at 
the front but they never grow to become dominating. 

In conclusion, we have presented experimental and theoretical
studies of the growth morphologies of colonies of the yeast 
{\em Pichia membranaefaciens}. As a function
of a control parameter, which in the experimental case is the
thickness of the agarose plate, very different morphologies 
are observed. These range from a ``cellular'' front with 
penetrating instabilities for a large agarose plate thickness, over
a morphology with deep grooves, to a compact colony with tiny
instabilities in the front. Using an analogy to 
directional solidification we introduce a coupling
between a diffusion equation for the products of the metabolism,
and the equation governing the propagation of the front.
We believe that the model could be applied at other parameter
ranges than utilized here, and that the model can be used to extrapolate 
and analyse morphologies at similar conditions for other types of
growing colonies. 


\clearpage
\begin{figure}
\caption[x]{
Time series showing the growth of {\em Pichia membranaefaciens} at 
$T=30^\circ {\rm C}$ and 100\% relative humidity. 
In (a) growth is directed from center and 
outwards, in (b) and (c) growth is directed upwards. 
(a) On solid agarose: 1/2 strength YPD, 3\,mm 0.8\,\% agarose film, 
    in 5-hour intervals. 
(c) Replenishing setup: 2.6\,mm, 1/2 strength YPD, 
    0.8\,\% agarose film, 
    1/16 strength replenished YPD, in 1-hour intervals.
(b) Replenishing setup: 1.0\,mm, 1/2 strength YPD, 
    0.8\,\% agarose film, 
    1/4 strength replenished YPD, 1-hour intervals.
\label{fig:1}
}
\end{figure}
\begin{figure}
\caption[x]{
Schematic view of the experimental setup with replenishing. 
An agarose film of thickness 0.3 - 2\,mm is suspended on a membrane 
over a continuously replenished and stirred YPD solution. 
With this setup, no retardation of the growth is observed. 
}
\label{fig:experiment1}
\end{figure}
\end{document}